\documentclass[aps,twocolumn,showpacs,pra]{revtex4}
\usepackage{mathrsfs}
\usepackage{graphicx}
\usepackage{dcolumn}
\usepackage{bm}
\usepackage{amsmath}
\usepackage{amsfonts}
\usepackage{txfonts}
\newtheorem{thm}{Theorem}
\newtheorem{corollary}{Corollary}

\begin{document}

\title{Dynamical decoupling for a qubit in telegraph-like noises}

\author{Ke Chen}

\author{Ren-Bao Liu}

\email{rbliu@cuhk.edu.hk}

\affiliation{Department of Physics, The Chinese University of Hong Hong, Shatin,
New Territories, Hong Kong, China}
\begin{abstract}
Based on the stochastic theory developed by Kubo and Anderson, we
present an exact result of the decoherence function of a qubit in
telegraph-like noises under dynamical decoupling control. We prove
that for telegraph-like noises, the decoherence can be suppressed
at most to the third order of the time and the periodic
Carr-Purcell-Merboom-Gill sequences are the most efficient
scheme in protecting the qubit coherence in the short-time limit.
\end{abstract}



\maketitle

\section{Introduction}

Dynamical decoupling (DD)
is a standard technique to suppress the spin decoherence,
with a long tradition in magnetic resonance spectroscopy~\cite{haeberlen1976hrn,Slichter1992}.
Nowadays, DD is important in areas like quantum information processing, since
prolonging the qubit coherence is a fundamental requirement to carry
out effective operations on quantum states. The basic idea of DD is
using a sequence of control pulses that frequently flip the spins
to average out effects of random environmental field.
DD is originated from Hahn's first spin echo experiment in 1950~\cite{Hahn:SpinEcho}.
After Hahn's work, more complex pulse sequences were introduced, among
which a most famous example was the periodic Carr-Purcell-Merboom-Gill sequence
(CPMG)~\cite{haeberlen1976hrn,Carr1954_CP},
which was initially widely used in magnetic resonance spectroscopy
and, in recent years, was introduced to protect the qubit coherence
in quantum information processing~\cite{ViolaLloyd,Ban1998,Zanardi199977,Viola1999}.
Concatenated DD (CDD)~\cite{Khodjasteh2005_PRL,Khodjasteh:2007PRA,Santos,Yao2007_RestoreCoherence,Liu2007,WitzelDasSarma}
is of special interest, since it recursively constructs pulse sequences
to eliminate qubit decoherence to an arbitrary order of precision
in the short-time expansion. The pulse number in CDD, however,
exponentially increases. The first optimal pulse sequence in terms of the pulse
number was introduced by Uhrig~\cite{uhrig:100504}, which was a non-periodic DD.
Uhrig's DD (UDD) was later shown to be universal~\cite{lee:160505,Yang&Liu},
in the sense that the leading $n$ orders of the time expansion of the decoherence are
eliminated by using $n$ pulses for general finite quantum systems, i.e.,  systems with
hard high-frequency cutoff in noise spectra. Recent experiments
achieved remarkable progresses in prolonging the coherence via DD
schemes~\cite{Morton2006,Biercuk09NatureUDD,DuJiangFeng09NatureUDD}.

An interesting question is how different DD schemes perform in suppressing spin or qubit decoherence caused by classical non-Gaussian noises, such as multi-state
telegraph-like noises. For Gaussian noises, which are fully characterized
by their second-order correlation functions, the DD control can be readily formulated as the integration of
noise spectra modulated by a filtering function determined by the DD sequence~\cite{cywinski:174509}.
For general non-Gaussian noises, however, such a formalism is not available and one has to rely
on cumulant expansion and/or Gaussian approximation to numerically solve the problem.
In Ref.~\cite{cywinski:174509}, numerical calculation shows that CPMG is practically better than CDD and UDD
for a telegraph noise. Later in Ref.~\cite{PasiniUhrig2010PRA},
numerical search for optimal DD sequences finds solutions close to the CPMG for noises with soft cutoffs
and to UDD for noises with hard cutoffs.
These research indicates that CPMG may be optimal in combating certain non-Gaussian noises, but
more research is still needed to reach the conclusive results.

In this paper, we present an exact result of the decoherence function under DD control,
based on a stochastic theory~\cite{Anderson,KuboNote,KuboStochastic}
developed in 1950s mainly by Kubo and Anderson, when they studied
the line shape of nuclear magnetic resonance (NMR) spectra. We prove
that for a general multi-state telegraph-like noise, any DD schemes cannot
fully eliminate the third order term in the short-time
expansion of the decoherence, i.e., the decoherence function is at least of $\mathcal{O}(t^{3})$,
and among all possible DD schemes, CPMG is the most optimal in suppressing the qubit decoherence in
the short-time limit. Apart from the theoretical importance, these results are relevant to the
spin decoherence problem in real systems such as spins in Si/SiO$_2$
interfaces, where the noises from coupling with dangling bonds can be
approximately represented by a discrete multi-state
telegraph-like noise~\cite{deSousa}.

The paper is organized as follows. In section \ref{sec:Method-to-obtain},
we obtain the exact expression for the decoherence function under arbitrary DD
control of a qubit in telegraph-like noises. In section \ref{sec:Time-expansion-and},
we expand the decoherence function, solve the optimization problem
and prove that CPMG is the global minimum solution.

\section{Exact decoherence function}
\label{sec:Method-to-obtain}

We consider a qubit (spin-1/2) under an external field and a random field $w(t)$. The Hamiltonian
 (in the rotating reference frame in which the external field is transformed to be zero) is
\begin{equation}
H=S_{z}w(t),
\end{equation}
where ${S}_{x/y/z}$ is the spin operator along the $x/y/z$ direction.
We assume $w(t)$ is a multi-state telegraph-like noise, i.e., it
jumps suddenly and randomly among a set of discrete values $\{w_j\}$,
with a transition rate matrix $\Gamma$, in the form
\begin{equation}
\frac{d}{dt}Y_j(t)=\Gamma_{jj'} Y_{j'}(t),
\end{equation}
where $Y_j(t)$ is the probability for $w(t)=w_j$.
Such a noise model is widely used in describing various stochastic
physical process, such as the spectral diffusion of optical transitions
due to atom collisions~\cite{BermanBrewer}.

The spin coherence is characterized by the transverse polarization
$x\equiv S_{x}+iS_{y}$.
The equation of motion for $x$ in the form of a so-called Kubo oscillator is
\begin{equation}
\dot{x}(t)=iw(t)x(t).
\label{eq:motion}
\end{equation}
The ensemble average $\left\langle x(t)\right\rangle\equiv \int x(t) P[w(t)]D[w(t)]$ over a distribution of
the random field $P[w(t)]$ gives the transverse polarization of the spin.
Defining $Y_j(\varphi,t)$ as the
probability density for finding $x(t)=e^{i\varphi}$ and
$w(t)=w_{j}$, we can write the ensemble averages
$\left\langle wx\right\rangle $ and $\left\langle x\right\rangle $ as
\begin{subequations}
\begin{align}
\left\langle wx\right\rangle & = \sum_{j}\int_{0}^{2\pi}e^{i\varphi}w_{j}Y_j(\varphi,t)d\varphi, \\
\left\langle x\right\rangle & =\sum_{j}\int_{0}^{2\pi}e^{i\varphi}Y_j(\varphi,t)d\varphi.
\end{align}
\end{subequations}
The probability density function satisfies the stochastic Liouville equation~\cite{KuboNote,KuboStochastic}
\begin{equation}
\frac{\partial}{\partial t}Y_j(\varphi,t)=\sum_{j'}\Gamma_{jj'}Y_{j'}(\varphi,t)-w_{j}\frac{\partial}{\partial\varphi}Y_j(\varphi,t),
\label{eq:stochastic eq}
\end{equation}
which contains both the sudden jumps of the random field and the precession of the spin polarization.
By multiplying $e^{i\varphi}$ to both sides of Eq.~(\ref{eq:stochastic eq}) and integration over the phase angle, we obtain
\begin{equation}
\frac{d}{dt}y(t)=(\Gamma+iW)y(t),
\label{Eq:evolution}
\end{equation}
where $y_{j}(t)\equiv\int_{0}^{2\pi}e^{i\varphi}Y_j(\varphi,t)d\varphi$,
$W_{jj'}=\delta_{jj'} w_{j'}$.
The spin polarization is $\left\langle x(t)\right\rangle =\sum_{j}y_{j}(t)$.

Under DD control, the spin is subjected to a sequence of flip operations between the $+z$ and $-z$ directions.
In this paper, for the sake of simplicity, we assume the case of ideal $\pi$ pulses.
Equivalently, the flip control can be transformed to the flip of the field $w(t)$
in the reference frame rested on the flipped spin. Thus, we have the controlled Liouville equation as
\begin{equation}
\frac{d}{dt}y(t)=\left[\Gamma+iWf(t)\right]y(t),
\label{Eq:DDevolution}
\end{equation}
where $f(t)$ is a step-like function jumping between $+1$ and $-1$ every time a control pulse is applied.
The exact solution of the decoherence function under
an $N$-pulse DD control is
\begin{align}
\left\langle x(t)\right\rangle & =  \sum_{j} \left[e^{\left[\Gamma+(-1)^{N}iW\right]a_{N+1}t}\cdots
e^{\left[\Gamma-iW\right]a_{2}t}e^{\left[\Gamma+iW\right]a_{1}t}y(0)\right]_{j},
 \label{Eq:exact}
 \end{align}
where $y(0)$ is the initial probability distribution of the random force,
and $0<a_{n}t<1$ is the interval between the $n$th and $(n-1)$th pulses with $\sum a_n=1$.

\section{\label{sec:Time-expansion-and}optimal dynamical decoupling}

To achieve a certain order of DD, we need to find the solution of the set of
(normalized) pulse intervals $\{a_i\}$ to make the decoherence function
$\langle x(t)\rangle$ equal to unity up to an error of a certain order
of $t$ in the short-time expansion. For this purpose, we expand the exponential functions
in Eq.~(\ref{Eq:exact}) into Taylor's series of $t$. The zeroth order term contains neither $W$ nor $\Gamma$, and is explicitly
$\sum_{j}y_{j}(0)=1$ (the sum probability must be unity).

The higher order terms can be classified by the ordering of the matrices $W$ and $\Gamma$.
A few rules can be established to significantly simplify the expansion.
First, any term starting with $\Gamma$ must vanish, because of
the probability conservation condition
\begin{equation}
\sum_j\Gamma_{jk}=0.
\end{equation}
Second, any term ending with $\Gamma$ must vanish since
\begin{equation}
\Gamma y(0)=0,
\end{equation}
for the random force distribution is stationary.
Third, the terms of the same order and containing no $\Gamma$ sum to zero.
This is because if $\Gamma$ is set to zero (corresponding to the static inhomogeneous broadening condition), the decoherence function
becomes
\begin{equation}
\left\langle x(t)\right\rangle =\sum_{j}\left[e^{iW(a_{1}-a_{2}+...+(-1)^{N}a_{N+1})t}y(0)\right]_{j}=1,
\end{equation}
under the echo condition $a_{1}-a_{2}+...+(-1)^{N}a_{N+1}=0$.
Using the three rules above, the only non-vanishing
term of the decoherence up to the  third order must
have the form of $G_N(a_1,a_2,\ldots,a_N)W\Gamma Wy(0)$.
To minimize the decoherence in the third order, we just need to minimize the
coefficient $G_N(a_1,a_2,\ldots,a_N)$.

With the conjecture that CPMG could be an optimal solution (as in the case of
two-pulse control for Gaussian noises with hard cutoffs), we write the coefficient $G_N(a_1,a_2,\ldots,a_N)$
as a function of the deviations of the pulse positions from the CPMG timing,
\begin{equation}
\beta_n\equiv \alpha_n-\frac{2n-1}{2N},
\end{equation}
where $\alpha_nt$ is the position of the $n$th control pulse (i.e., $a_n\equiv \alpha_n-\alpha_{n-1}$).
The echo condition is $\beta_{1}-\beta_{2}+\cdots+(-1)^{N+1}\beta_{N}=0$,
and another constraint is $1>\alpha_{N}>\alpha_{N-1}>\cdots>\alpha_{1}>0$.
These two conditions define the physical boundary for an $N$-pulse sequence.
Then the coefficient of the third order term is
\begin{widetext}
\begin{align}
G_N(\beta) = &  \frac{1}{12N^{2}}
 +\frac{1}{N}\left\{\left[-\beta_{N}+2\beta_{N-1}-2\beta_{N-2}+...+(-1)^{N}\beta_{1}\right]^{2}
  +\left[-\beta_{N-1}+2\beta_{N-2}+\cdots +(-1)^{N-1}\beta_{1}\right]^{2}
  +\cdots +\left[-\beta_{1}\right]^{2}\right\}
\nonumber \\
 & + \left\{ 2\beta_{N}^{2}\left[-\beta_{N}+2\beta_{N-1}-2\beta_{N-2}+...+(-1)^{N}\beta_{1}\right]
   +2\beta_{N-1}^{2}\left[-\beta_{N-1}+2\beta_{N-2}+\cdots+(-1)^{N-1}\beta_{1}\right]
 +\cdots
 +2\beta_{1}^{2}\left(-\beta_{1}\right)\right\}
 \nonumber \\
\equiv & \frac{1}{12N^{2}} +h_N+g_N,
\label{eq:def}
\end{align}
\end{widetext}
where $h_N$ denotes the second-order term,
and $g_N$ the third-order term.
The first order term vanishes because of the echo condition.
Since the second order is always positive ($h_N>0$), CPMG ($\beta=0$) must be at least a local minimum
solution of the decoherence function.

If CPMG is not the global minimum, we can find a set of $\{\beta\ne 0\}$ so that
$h_N(\beta)+g_N(\beta)<0$. Since $h_N(\beta)>0$, we must have $g_N(\beta)<0$.
Using a real number $\lambda$ to scale the deviation, we get a function of the scaling factor as 
\begin{equation}
f_N(\lambda)\equiv g_N(\lambda\beta)+h_N(\lambda\beta)=\lambda^{3}g_N(\beta)+\lambda^{2}h_N(\beta).
\end{equation}
The function $f_N(\lambda)$ monotonically decreases with $\lambda$ for $\lambda\ge 1$.
Thus when $\lambda$ is increased from 1, the DD sequence suppresses the decoherence better and better.
But $\lambda$ cannot be infinitely increased under the physical conditions.
When $\lambda$ is increased to a boundary value $\lambda_B$, the deviation $\lambda\beta$
will reach the physical boundary, at which either two adjacent pulses coincide (and become a
null operation) or a pulse reaches the boundary time
at $0$ or $t$. This means a new pulse sequence with fewer pulses is obtained.
Suppose the new pulse sequence has $N'$ pulses (with $N'<N$), the coefficient
can be written as
\begin{equation}
\frac{1}{12N^{2}}+f_{N}(\lambda_B)=\frac{1}{12{N'}^{2}}+g_{N'}(\beta')+h_{N'}(\beta'),
\end{equation}
where $\beta'$ is the deviation from the $N'$-pulse CPMG, and $g_{N'}$ and $h_{N'}$ are defined
as in Eq.~(\ref{eq:def}). Obviously,
\begin{equation}
g_{N'}(\beta')+h_{N'}(\beta')<0.
\end{equation}
Then following the same procedure as in the $N$-pulse sequence,
we can do the scaling $\lambda'\beta'$ again to find pulse sequences
performing better in suppressing the decoherence by
increasing $\lambda'$ from 1 to a new boundary value.
Then we would find a new sequence which performs better with fewer pulses than $N'$.
So on and so forth, we will have to conclude that the one-pulse sequence (Hahn echo)
is the optimal solution among all sequences (including the multiple-pulse ones),
which can be easily checked to be wrong by comparing the performance of
the two-pulse CPMG and the Hahn echo. Thus, to avoid contradictions, we conclude with the
following theorem:
\begin{thm}
For an arbitrary multi-state telegraph-like noise, CPMG is the global minimum solution
of the decoherence function among all DD sequences of the same number of pulses.
\end{thm}
A corollary which can be directly derived from Eq.~(\ref{eq:def}) is
\begin{corollary}
The decoherence of a qubit in a multi-state telegraph-like noise
can not be suppressed by DD beyond the third order of the
short-time expansion.
\end{corollary}
This conclusion is consistent with previous numerical solutions in Ref.~\cite{PasiniUhrig2010PRA},
which give pulse sequences close to CPMG for boson baths with power-law high-frequency cutoffs.
Actually, the second order correlation function of the telegraph-like noise
\begin{equation}
\langle w(t)w(t')\rangle=
 \sum_j\left[We^{-\Gamma (t-t')}Wy(0)\right]_j,
\end{equation}
has the form of exponential functions. The Fourier transformation to the frequency domain then
has a power-law decay profile in the high frequency end.
Similarly, the higher order correlation functions (which in general cannot be factorized into the second order
correlation functions as for Gaussian noises) also have the form of exponential functions.
Such exponential function form of the correlation functions results from the sudden jumps in the
telegraph-like noises, which, in the physical nature, is induced by instantaneous
(Markovian) collisions in the bath. It is the existence of a very short timescale in the
system (the collision memory time), or in other words, a very large energy scale,  that
limits the performance of any DD schemes to the third order of precision.

\section{Conclusions}

In summary, we have derived an exact expression for the decoherence
function of a qubit in arbitrary multi-state telegraph-like noises,
based on the stochastic theory.
We prove that CPMG is the globally optimal solutions among all possible dynamical decoupling
sequences of the same number of pulses to suppress the decoherence in the short time limit.
Because of the instantaneous random jumps in the noises, the decoherence
cannot be eliminated beyond the third order of the short time.

\begin{acknowledgments}
This work was supported by Hong Kong RGC CUHK/402209.
\end{acknowledgments}



\end{document}